\documentclass[twocolumn,preprintnumbers,amsmath,amssymb,showkeys]{revtex4}
\usepackage{graphicx}% Include figure files
\usepackage{dcolumn}% Align table columns on decimal point
\usepackage{bm}% bold math

\begin{document}

\preprint{}

\title{Observation of the Goos-H\"{a}nchen shift in graphene via weak measurements}% Force line breaks with \\
\author{Shizhen Chen}
%\altaffiliation[ ]{}%Lines break automatically or can be forced with \\
\author{Chengquan Mi}
\author{Liang Cai}
\author{Mengxia Liu}
\author{Hailu Luo}\email{hailuluo@hnu.edu.cn}
\author{Shuangchun Wen}
%%\email{Second.Author@institution.edu}
\affiliation{Laboratory for spin photonics, School of
 Physics and Microelectronics Science, Hunan University, Changsha 410082,China}
\date{\today}% It is always \today, today,

\begin{abstract}
We report the observation of the Goos-H\"{a}nchen effect in graphene
via a weak value amplification scheme. We demonstrate that the
amplified Goos-H\"{a}nchen shift in weak measurements is sensitive
to the variation of graphene layers. Combining the Goos-H\"{a}nchen
effect with weak measurements may provide important applications in
characterizing the parameters of graphene.
\end{abstract}

%\pacs{42.25.-p, 42.79.-e, 41.20.Jb}% PACS, the Physics and Astronomy

%Use showkeys class option if keyword
\maketitle

The behavior of plane wave in reflection can be simply
predicted by geometrical optics. However, for the bounded beam of
light, it may undergo extra shifts due to the occurrence of
diffractive corrections. Such shifts are known as the
Goos-H\"{a}nchen (GH)~\cite{Goos1947} and the Imbert-Fedorov
~\cite{Imbert1972,Fedorov1955} shifts according to the directions
parallel and perpendicular to the plane of incidence, respectively.
In recent years, the research of the GH shift is still active
although it was discovered more than 60 years ago. The corresponding
studies reach not only to the investigation of the inherent physics
behind this
phenomenon~\cite{Aiello2008,Dennis2012,Toppel2013,Bliokh2013,Ornigotti2015},
but also to the behavior of the shift at various reflecting
surfaces~\cite{Gilles2002,Lai2002,Qing2004,Yin2006,He2006,Merano2007},
especially the theoretical works of the GH shift in reflection from
a graphene-dielectric interface. The system with graphene is very interesting for the observation of beam shifts due to its flexible optical properties. The Fresnel reflection coefficients with the existence of graphene become
different~\cite{Grosche2015,Hermosa2016}, and the behavior of the GH
shift is changed or tunable~\cite{Martinez2011,Cheng2014}. In particular, the GH shift on a substrate coated with graphene can be
quantized in an external magnetic field~\cite{Kort-Kamp2016}. Even though in a common environment,
a giant GH shift in graphene was observed recently~\cite{Li2014}.

In this paper, we investigate the GH shift in graphene under a total
internal reflection (TIR) condition. Like other beam shifts, the GH is generally very small and difficult to be directly observed. Here we amplify it via a weak measurement technique to overcome this difficulty. Interestingly, we find that the shift amplified
by a weak value scheme is sensitive to the graphene layers. Weak measurements are an important and
convenient approach that has reached fruitful achievements for
detecting light beam
shifts~\cite{Hosten2008,Dixon2009,Qin2009,Zhou2012I}. This novel
conception was first proposed in the context of quantum mechanics
and then has been extensively
studied~\cite{Aharonov1988,Duck1989,Kofman2012,Dressel2014}. With
the help of weak measurements, the GH shift in
TIR~\cite{Jayaswal2013} or in partial reflection~\cite{Goswami2016}
can be observed recently. In our work, the GH shift occurs in the regime of TIR (the reflected
intensity is equal to incident intensity), and the success rate of
the postselection is still large due to no requirement of a very
large weak value. That is, the final output intensity is still strong and measured
data are stable. Our result suggests that this technique may become an alternative way to effectively and conveniently identify layers of few-layer graphene.

\begin{figure}
\centerline{\includegraphics[width=8cm]{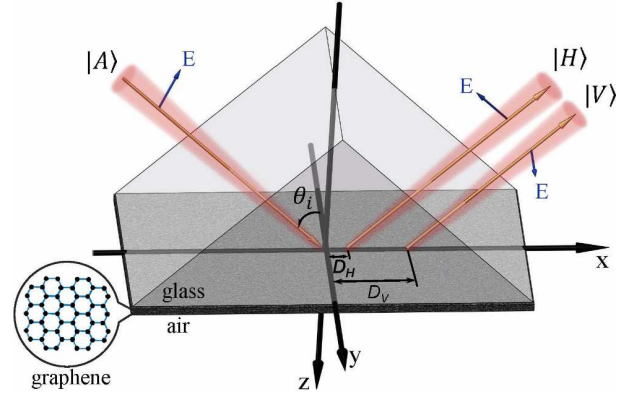}}
\caption{\label{Fig1}(color online) Schematic illustrating the GH shift in
graphene under TIR. A $45^\circ$ linearly polarized beam labeled by
$|A\rangle$ hits a glass-graphene-air interface at an incident angle
$\theta_{i}$. Then the components $|H\rangle$ and $|V\rangle$
experience lateral small GH shifts $D_{H}$ and $D_{V}$,
respectively.}
\end{figure}

\begin{figure}
\centerline{\includegraphics[width=8cm]{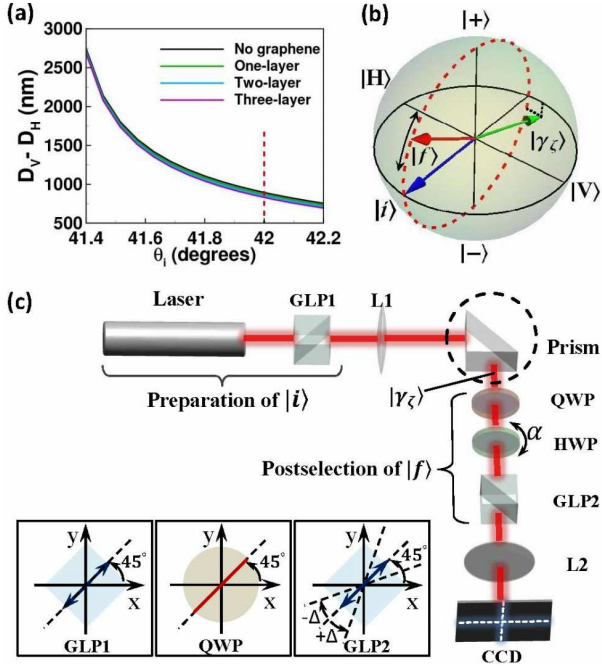}}
\caption{\label{Fig2} (Color online) (a) ($D_{V}$-$D_{H}$) as a function of
incident angle for different graphene layers. The red dashed line
indicates the incident angle ($42^\circ$) in our experiment, the
angle of total reflection is $41.3^\circ$. (b) Representation on the
Bloch sphere of the states $|i\rangle$, $|\gamma_{\zeta}\rangle$,
and $|f\rangle$. $|i\rangle$ is the incident state and $|f\rangle$
is the postselected state. $|\gamma_{\zeta}\rangle$ is the
preselected state after reflecting at the graphene-dielectric
interface. The state $|\gamma_{\zeta}\rangle$ is off to be
antiparallel to $|i\rangle$ in two angular directions depending on
the graphene layer numbers. The red dashed circle represents the
trajectory of $|f\rangle$ when we rotate the half-wave plate
(HWP). (c) Experimental setup: a Gaussian beam generated by a He-Ne
laser ($632.8\mathrm{nm}$, Thorlabs HNL210L-EC). GLP1 and GLP2, Glan
Laser polarizers; QWP, quarter-wave plate; L1 and L2, lenses with
focal length 125mm and 250mm respectively. The beam waist after L1 is 71.25 $\mu$m. The data are detected by a CCD camera
(Coherent LaserCam HR). Insets show the rotations of GLP1, QWP, and
GLP2.}
\end{figure}

Consider a reflection system shown in Fig.~\ref{Fig1}. An incoming
beam is at the $45^\circ$ linearly polarization state $|A\rangle$.
Under a TIR condition, the reflected light consists of two beams,
one is the horizontal ($H$) component $|H\rangle$ displaced by
$D_{H}$ and the other is the vertical ($V$) component $|V\rangle$
displaced by $D_{V}$. We only concern about the difference between
$D_{V}$ and $D_{H}$ because it is the measured variable in our weak
measurements scheme. To calculate the GH shift ($D_{V}- D_{H}$) for
different graphene layers, the obtainment of the reflection
coefficient in the graphene-dielectric interface is important. The
reflection coefficient here is given by~\cite{Zhou2012II,Li2014}
\begin{equation}
r_{A}=\frac{R_{A}+R_{A}^{'}\exp\left(2idk_{gz}
\right)}{1+R_{A}R_{A}^{'}\exp\left(2idk_{gz} \right)}\label{asi}.
\end{equation}
Here, $R_{A}$ and $R_{A}^{'}$ are the Fresnel reflection
coefficients in the glass-graphene and graphene-air interfaces,
respectively, $A\in\{H,V\}$. $k_{gz}$ is the component of wave
vector $k_{0}$ in graphene alone $z$ direction. $d=\zeta\Delta d$ is
the thickness of the graphene film with $\zeta$ and $\Delta
d=0.34$nm representing the layer numbers and the thickness of single
layer graphene, respectively. In TIR, the reflection coefficient
$r_{A}$ is complex and can be written as
$r_{A}=|r_{A}|\exp(i\varphi_{A})$. Neglecting the small angular
shift due to the slow variation of $|r_{A}|$~\cite{Grosche2015}, the
GH shifts for $H$ and $V$ components are spatial and can simply form
as
\begin{equation}
D_{A}=-\frac{1}{n
k_{0}}\frac{\partial\varphi_{A}}{\partial\theta_{i}}\label{hj}.
\end{equation}
Here, $n=1.515$ is the refractive index of prism. From
Eq.~(\ref{hj}), we plot the curves of the GH shift as a function of
incident angle for different graphene layers with the refractive
indexes ($3+1.149i$) of graphene, as shown in Fig.~\ref{Fig2}(a).
The shift decreases with increasing numbers of layer, but the
decrement is tiny and all shifts are small.

To amplify these small shifts, the weak measurements are employed
and the experimental setup is plotted in Fig.~\ref{Fig2}(c). This setup is similar to that prescribed in~\cite{Jayaswal2013}. In a
schema of weak measurements, the preselected state, postselected
state, and weak coupling between the system and pointer are three key elements. The corresponding elements in our scheme will be clear in the following.

\begin{figure}
\centerline{\includegraphics[width=8cm]{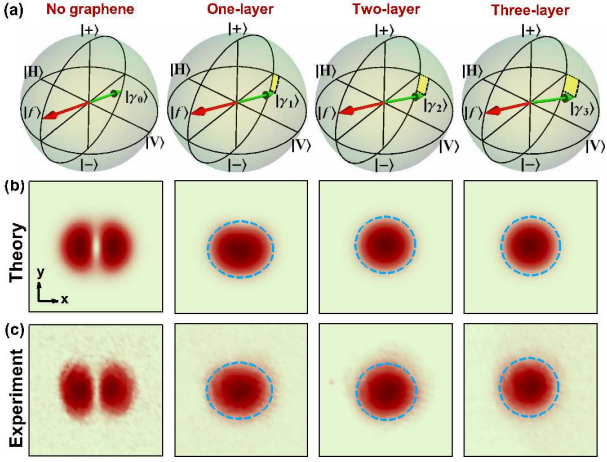}}
\caption{\label{Fig3} (Color online) (a) Representation on the Bloch sphere of the
state $|\gamma_{\zeta}\rangle$ and $|f\rangle$ when the tunable
state $|f\rangle$ is closest orthogonal to $|\gamma_{\zeta}\rangle$
(reading intensity on CCD becomes minimum). The yellow area
indicates the deviation to the circle of $|f\rangle$. (b) The
theoretical minimum intensity by adjusting HWP. (c) The
corresponding minimum intensity we read out from CCD. Note that the
first, second, third, and fourth columns correspond to the cases of
no, one-layer, two-layer, and three-layer graphene, respectively.}
\end{figure}

We first discuss the preselected state which is delicate here due to
the effect of reflection coefficients. In the experiment, we set the
optical axis of GLP1 to $45^\circ$ to project the incident
polarization state on $|i\rangle=(1/\sqrt{2},1/\sqrt{2})$. And this
polarization state changes to $|\gamma_{\zeta}\rangle=F|i\rangle$
when the light is reflected at the interface. $F$ is the reflection
matrix~\cite{Gotte2013}
\begin{eqnarray}
F =\left[
\begin{array}{cc}
-r_{H}&0 \\
0 & r_{V}
\end{array}
\right],
\end{eqnarray}
of which action is also a part of the preselection process, and the
preselected state for our weak measurements scheme is
$|\gamma_{\zeta}\rangle$. Note that the state $|\gamma_{\zeta}\rangle$ is an elliptical polarization
state which is different for one, two, and three graphene layers. Then we consider the weak coupling. The
tiny GH effect is regarded as a weak measuring process here, as labeled
by the dashed circle in Fig.~\ref{Fig2}(c). In the language of
quantum mechanics, this effect described by operator
is~\cite{Toppel2013,Jayaswal2014I,Goswami2016}
\begin{eqnarray}
\hat{\rm GH} =\left[
\begin{array}{cc}
D_{H}&0 \\
0 & D_{V}
\end{array}
\right]\label{matrixr}.
\end{eqnarray}
We next analyze the postselection, which can be realized by the
combination of QWP, HWP, and GLP2. In the experiment, the optical
axis of QWP is fixed to $45^\circ$  from the $x$ axis, and the
rotation angle of HWP is $\alpha$. These settings described by the
Jones matrices are
\begin{gather}
\rm QWP=\frac{1}{\sqrt{2}}\left[
\begin{array}{cc}
1&-i \\
-i & 1
\end{array}
\right] \nonumber\\ \rm HWP=\left[
\begin{array}{cc}
\cos(2\alpha)&\sin(2\alpha) \\
\sin(2\alpha) & -\cos(2\alpha)
\end{array}
\right]\label{as}.
\end{gather}
The optical axis of GLP2 is ($45^\circ\pm\Delta$) to project on the
state
\begin{gather}
\langle\rm GLP2|=\left[
\begin{array}{cc}
\cos(45^\circ\pm\Delta)&\sin(45^\circ\pm\Delta)
\end{array}
\right]\label{bc}.
\end{gather}
Putting Eqs.~(\ref{as}) and ~(\ref{bc}) together, we obtain the
postselected state as
\begin{gather}
\langle f|=\left[
\begin{array}{cc}
e^{i(2\alpha\mp\Delta)}&-e^{-i(2\alpha\mp\Delta)}
\end{array}
\right]\label{kj}.
\end{gather}
For convenience, we represent above states $|i\rangle$,
$|\gamma_{\zeta}\rangle$, and $|f\rangle$ on the Bloch sphere, as shown
in Fig.~\ref{Fig2}(b). One can see that the postselected state
$|f\rangle$ is limited on the red dashed circle by adjusting HWP [from Eq.~(\ref{kj})]. For the preselected state
$|\gamma_{\zeta}\rangle$, it exhibits a deviation
from the red dashed circle on the Bloch sphere in Fig.~\ref{Fig2}(b) [see Fig.~\ref{Fig3}(a) for the cases of different layers]. The reason for this deviation is that $|r_{H}|\neq|r_{V}|$ and the difference between $|r_{H}|$ and
$|r_{V}|$ increases with the increased layers of graphene.

With the preselected and postselected states discussed above, the
weak value in our weak measurements can be obtained as
\begin{equation}
\frac{\langle f|\hat{\rm GH}|\gamma_{\zeta}\rangle}{\langle
f|\gamma_{\zeta}\rangle}=\frac{1}{2}(D_{H}+D_{V})+\frac{A_{w}}{2}(D_{H}-D_{V})\label{weak},
\end{equation}
where $A_{w}=\langle
f|\hat{\sigma_{3}}|\gamma_{\zeta}\rangle/\langle
f|\gamma_{\zeta}\rangle$, and $\hat{\sigma_{3}}$ is the Pauli
matrix. Obviously, the weak value from Eq.~(\ref{weak}) is related
to $\zeta$, i.e., the layer of graphene. Here, $A_{w}$ in all cases
is a complex number except the one of $\zeta=0$, in which it is pure
imaginary, for the full study in different theory about this case one can see~\cite{Jayaswal2013}. Note that the GH shift in TIR is a
spatial shift (the angular GH shift is too small even in graphene),
and the imaginary part of $A_{w}$ can naturally convert the relevant
spatial shift ($D_{H}-D_{V}$) into an angular one. Therefore, in
order to obtain the centroid position $\langle x\rangle$ of the
beams, the propagation effect in all cases should be
considered~\cite{Hosten2008,Aiello2008}. Containing
Eq.~(\ref{weak}), $\langle x\rangle$ is obtained as
\begin{equation}
\frac{1}{2}[(D_{H}+D_{V})+\mathrm{Re}(A_{w})(D_{H}-D_{V})+\frac{z}{z_{r}}\mathrm{Im}(A_{w})(D_{H}-D_{V})]\label{fin},
\end{equation}
where $z_{r}$ is the Rayleigh range and $z$ is the propagation. In
fact, the shift we measure in the experiment is a relative position
of the beams on CCD for the postselected states with $\pm\Delta$.
Thus, the third term in Eq.~(\ref{fin}) is pivotal and the first term is irrelative. The second term
mainly results from the inequality between $|r_{H}|$ and $|r_{V}|$.

We now turn to the procedure for experimentally observing amplified
GH shift. The rotations of GLP1 and QWP are all fixed at $45^\circ$. We first set GLP2 to $45^\circ$, i.e., $\Delta=0$. Then we
rotate the HWP with an angle $\alpha$ and the state $|f\rangle$ will project on the red dashed circle in Fig.~\ref{Fig2}(b). Note
that the rotation $\alpha$ is different for different layers of graphene, but it is unnecessary to clarify it in the experiment.
We adjust the HWP until the output intensity on CCD becomes minimize, which indicates
that the postselected state $|f\rangle$ is closest
orthogonal to the preselected state $|\gamma_{\zeta}\rangle$. After minimizing the output intensity, we rotate the GLP2 first to ($45^\circ+\Delta$) and then to
($45^\circ-\Delta$) to measure the final amplified shift.

\begin{figure}
\centerline{\includegraphics[width=8cm]{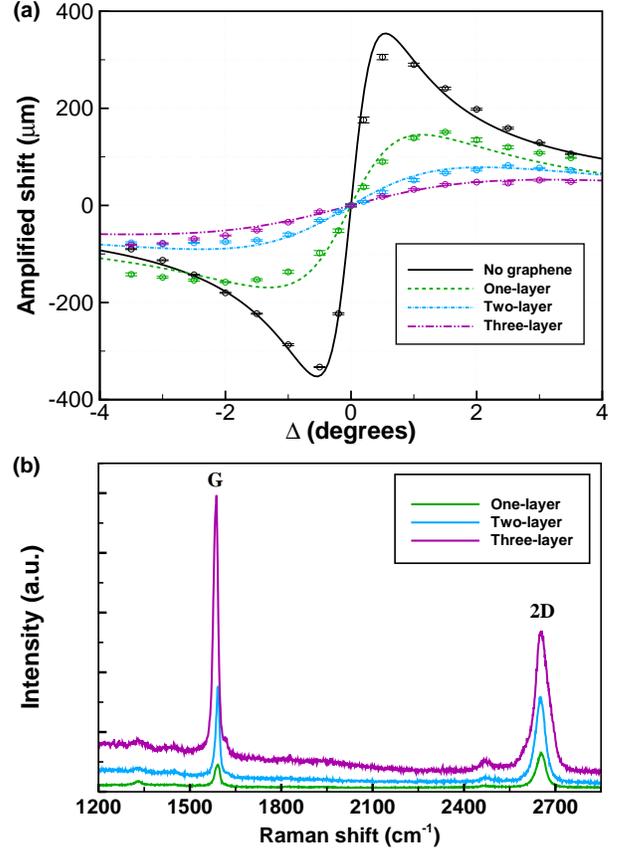}}
\caption{\label{Fig4} (Color online) (a) The amplified shift as the function of
angle $\Delta$ for different graphene layers. Experimental data are
shown as open dots with error bars. (b) Raman spectrum of our samples for one-layer, two-layer, and three-layer
graphene.}
\end{figure}

As discussed above, we use a quantum mechanical description to analyze the amplified GH shift in order to provide a good physical insight and simplify the analysis. In fact, the process for the weak measurement of GH effect can be described by using standard
wave optics~\cite{Jayaswal2013}, and the simulative minimum output intensity is illustrated in Fig.~\ref{Fig3}(b). We see that only in the case of no
graphene the minimum intensity exhibits double-peak profile, which is a common
distribution of minimum output intensity in weak
measurements~\cite{Duck1989}. This is because in this case the
state $|f\rangle$ is possible to be orthogonal to
$|\gamma_{0}\rangle$. For other cases with the existence of
graphene, the tunable postselected state can not be orthogonal to
the preselected state, and the nonorthogonal degree becomes larger
when the layers of graphene increase, leading to a smaller weak
value. As a result, the minimum intensity tends to a Gaussian form~\cite{Chen2015}. The corresponding minimum intensity we experimentally observe is shown as Fig.~\ref{Fig3}(c).

We measure the amplified GH shifts in no, one-layer, two-layer, and
three-layer graphene. The theoretical results of the amplified shift are briefly described by Eq.~(\ref{fin}). However, in order to avoid the approximate limits and get accurate values, the curves in Fig.~\ref{Fig4}(a) are given by a precise weak measurement theory~\cite{Kofman2012}. For a fixed angle $\Delta$, the
amplified shift decreases with increasing layers. Thus, we can conveniently determine the layers
of a sample at a special angle. In our experiment, each sample we fabricate is uniform layer and the size of sample is $1\mathrm{cm}\ast1\mathrm{cm}$.
In practice, we repeat the experiment of weak measurements several times in different measuring place for each sample
and the data are nearly same. To confirm the corresponding layers of graphene film, we provide their Raman spectra in Fig.~\ref{Fig4}(b).
The layers of each sample deduced from our data coincide well with the results from Raman spectra.

The measurability of very small displacements
is ultimately limited by the quantum noise of the
light, because enough photons need to be collected
to resolve the position of the field~\cite{Hosten2008}.
Note that another interesting beam
shift induced by photonic spin Hall effect can also be used to
identify graphene layers~\cite{Zhou2012II}. In that case, the experimental measurement was performed near Brewster angle.
Therefore, the experimental data are a little unstable due to a low reflection intensity near Brewster angle.
In present case, however, enough photons can be captured by the detector due to the TIR.
From the error bars in Fig.~\ref{Fig4}(a), we see that the data read from CCD are very stable.
We recently noted that a similar experimental setup can also be used to observe the angular and lateral GH shifts near the critical angle for TIR~\cite{Santana2016}.

In conclusion, we have experimentally observed the GH shift in
graphene via weak measurements. Theoretically, the variations of the
initial shifts for no, one-layer, two-layer, and three-layer
graphene are tiny. However, employing a weak value amplification
scheme, the amplified GH shift decreases with the increased layers
of graphene. This technique may be utilized to identify layers of few-layer graphene with the
stable data of this system. Our research is important for the
research of graphene in future and may offer the opportunity to
characterize the parameters of graphene with the help of weak measurements.

\begin{acknowledgements}
This research was supported by the National Natural Science Foundation
of China (Grants Nos. 11274106 and 11474089); Hunan Provincial Innovation
Foundation for Postgraduate (CX2016B099).
\end{acknowledgements}

\end{document}